\documentclass[manuscript]{aastex}
\usepackage{color}
\usepackage{amsmath}
\shorttitle{}
\shortauthors{Monje et al.}

\begin{document}


\title{Hydrogen Fluoride toward Luminous Nearby Galaxies: NGC~253 and NGC~4945}


\author{R. R. Monje} 
\affil{California Institute of Technology, 1200 E. California Blvd., Pasadena, CA  91125-4700, USA}
\email{raquel@caltech.edu}
\author{S. Lord}
\affil{Infrared Processing and Analysis Center, California Institute of Technology, MS 100-22, Pasadena, CA 91125, USA}
\author{E. Falgarone}
\affil{LERMA/LRA, Ecole Normale Sup\'erieure \& Observatoire de Paris, 24 rue Lhomond, 75005 Paris, France}
\author{D. C. Lis}
\affil{California Institute of Technology, 1200 E. California Blvd., Pasadena, CA  91125-4700, USA}
\author{D. A. Neufeld}
\affil{Department of Physics and Astronomy, Johns Hopkins University, 3400 North Charles Street, Baltimore, MD 21218, USA}
\author{T. G. Phillips} 
\affil{California Institute of Technology, 1200 E. California Blvd., Pasadena, CA  91125-4700, USA}
\author{and R. G\"{u}sten} 
\affil{Max-Planck Institut f\"{u}r Radioastronomie, Auf dem H\"{u}gel 69, 53121 Bonn, Germany}
%\email{aastex-help@aas.org}
%\affil{Space Telescope Science Institute, Baltimore, MD 21218}


%\altaffiltext{1}{Visiting Astronomer, Cerro Tololo Inter-American Observatory.
%CTIO is operated by AURA, Inc.\ under contract to the National Science
%\altaffiltext{2}{Society of Fellows, Harvard University.}
%\altaffiltext{3}{present address: Center for Astrophysics,
%    60 Garden Street, Cambridge, MA 02138}
%\altaffiltext{4}{Visiting Programmer, Space Telescope Science Institute}
%\altaffiltext{5}{Patron, Alonso's Bar and Grill}


\begin{abstract}

We present the detection of hydrogen fluoride, HF, in two luminous nearby galaxies NGC~253 and NGC~4945 using the Heterodyne Instrument for the Far-Infrared (HIFI) on board the {\it Herschel} Space Observatory. The HF line toward NGC~253 has a P-Cygni profile, while an asymmetric absorption profile is seen toward NGC~4945. The P-Cygni profile in NGC~253 suggests an outflow of molecular gas with a mass of {\it M}(H$_2$)$_{out}$~$\sim$ 1~$\times$~10$^7$~M$_\odot$ and an outflow rate as large as {\it \.{M}} $\sim$~6.4~M$_\odot$ yr$^{-1}$. In the case of NGC~4945, the axisymmetric velocity components in the HF line profile is compatible with the interpretation of a fast-rotating nuclear ring surrounding the nucleus and the presence of inflowing gas. The gas falls into the nucleus with an inflow rate of $\le$~1.2~M$_\odot$ yr$^{-1}$, inside a inner radius of $\le$ 200 pc. The gas accretion rate to the central AGN is much smaller, suggesting that the inflow can be triggering a nuclear starburst. From these results, the HF $J = 1-0$ line is seen to provide an important probe of the kinematics of absorbing material along the sight-line to nearby galaxies with bright dust continuum and a promising new tracer of molecular gas in high-redshift galaxies.

\end{abstract}


\section{Introduction}
%\textcolor{red}{Tom: Do you have a case for CO (or H I) where you can show what an infall with an AGN looks like, compared to a starburst?}

%\textcolor{red}{Darek: win you compute Gamma c or l, do you include the emission component as additional background?}

%Hydride molecules, compounds form with one or more hydrogen and any other element from the periodic table, are particularly important in astrochemistry why?: since they are 

{\it Herschel} Space Observatory \citep{Pil10}, and in particular its Heterodyne Instrument for Far-Infrared (HIFI) \citep{deG10} has facilitated observations at high-spectral resolution of interstellar hydride molecules - compounds with one or more hydrogen and one heavy element atom -, in the local universe (z $\sim$ 0). Most of the hydride molecule observations are inaccessible from ground-based telescopes because their high frequency rotational transitions are blocked by the large opacity of Earth's atmosphere. Two key results from {\it Herschel}/HIFI were the first detection of the \emph{fundamental} $J=1-0$ rotational transition of hydrogen fluoride (HF) at 1.232 THz (243 $\mu$m) and the discovery of HF's ubiquitous nature in the Milky Way. HF has not only been observed in almost every bright continuum source in the Galactic plane \citep{Phi10,Son10,Neu10,Mon11a}, but also in some nearby ultra luminous galaxies \citep{van10,Ran11}, establishing its importance outside the Milky Way as well.
%HF is form after the exothermal reaction of fluorine and H$_2$
%Fluorine can be found mainly in its neutral form in the diffuse ISM    
%HF chemistry: 
Despite Fluorine's (F) relative low abundance in the interstellar medium (ISM) (about four times lower than carbon), F plays an important role in the interstellar chemistry due to the unique thermochemistry of the reaction between F and molecular hydrogen (H$_2$). F is the only atom that reacts exothermically with H$_2$, to form the compound HF. Once formed, HF becomes the main reservoir of fluorine in the ISM, with a strong bond only destroyed by reactions with low abundance ions H$^{+}_{3}$, C$^{+}$ and He$^{+}$, or photodissociation, with an estimated photodissociation rate of 1.17~10$^{-10}$~s$^{-1}$. This unusual stability allows the build up of large amounts of HF in the ISM, which has now been confirmed by {\it Herschel}. 

%\textcolor{red}{Hydrogen fluoride (HF) is expected to be the main reservoir of fluorine (F) in the interstellar medium (ISM) because of its unique thermochemistry. Fluorine, predominantly neutral in the diffuse ISM, reacts exothermically with H$_{2}$, the dominant constituent of molecular clouds, forming HF. Once the H$_{2}$ abundance becomes appreciable, HF becomes the dominant reservoir of fluorine and is destroyed only slowly by photodissociation, with an estimated photodissociation rate of 1.17~10$^{-10}$~s$^{-1}$ for the mean {\bf interstellar} radiation field, and by reactions with He$^{+}$, H$^{+}_{3}$, and C$^{+}$. }

Theoretical models from \cite{Neu05} originally predicted that the HF abundance in diffuse clouds of small extinction ({\it A$_V$} $<$ 0.5), could be larger than that of the carbon monoxide (CO) - the main tracer of molecular hydrogen in the submillimeter range- despite the lower abundance of gas-phase fluorine compared to that of carbon. Furthermore, the models also showed that HF may be a more reliable tracer of H$_2$ than CO, since it presents a constant HF/H$_2$ ratio (see Figure 6 in Neufeld et al. 2005) with respect to the total visual extinction through the cloud, while the CO/H$_2$ ratio decreases in clouds of small {\it A$_V$}. \\
Thus, the \cite{Neu05}  model predicted that the ground state rotational transition line of HF $J=1-0$ would yield an extremely sensitive probe of the diffuse molecular gas along the line-of-sight to background far-infrared continuum sources and a potential valuable surrogate for molecular hydrogen. 
%HF thus becomes the main reservoir of fluorine in the interstellar medium. 
%\textcolor{red}{To date, such material has been studied in the submillimeter range primarily by means of carbon monoxide (CO) rotational emission lines. However, \cite{Neu05} show in their models that, in diffuse clouds of small extinction, the predicted HF abundance can actually exceed that of CO, even though the gas-phase fluorine abundance is 4 orders of magnitude smaller than that of carbon. HF is a more reliable tracer than CO because the HF/H$_{2}$ ratio is constant \citep[see Figure~6 in][]{Neu05}, whereas CO/H$_{2}$ drops in clouds of small {\it A$_{\rm v}$}.}
The high Einstein coefficient for spontaneous emission coefficient of the HF $J=1-0$ line, {\it A$_{ \rm 10}$}~=~2.42~10$^{-2}$~s$^{-1}$, results in simple excitation, with very high critical density ($\sim$ 10$^9$ cm$^{-3}$), at which collisional processes become important for the excitation.
Thus, under typical conditions characteristic of the diffuse or even dense ISM, the HF molecules are mainly in the ground rotational state.
 As a result of its large {\it A$_{10}$} coefficient the HF $J=1-0$ line has also been observed primarily in absorption. Only an extremely dense region or a strong radiation field could generate enough excitation to yield an HF feature with a positive frequency-integrated flux. 
%\textcolor{red}{ Thus, essentially all HF molecules are in the ground rotational state under typical conditions characteristic of the diffuse or even dense ISM.}
%{\bf Observations in the local universe of the fundamental HF $J=1-0$ rotational transition at 1.232 THz (243 $\mu$m) has been achievable for 
%the first time with the use of the Heterodyne Instrument for Far-Infrared (HIFI) \citep{deG10} on board the {\it Herschel} Space Observatory \citep{Pil10}.}  
%\textcolor{blue}{The Heterodyne Instrument for Far-Infrared (HIFI) \citep{deG10} on board the {\it Herschel} Space Observatory \citep{Pil10} has allowed for the first time the detection of the \emph{fundamental} $J=1-0$ rotational transition of HF at 1.232 THz (243 $\mu$m) in the local universe. HF has proven to be a ubiquitous tracer of molecular gas in the ISM of the Milky Way, being detected in environments as diverse as Orion~KL, OMC-1 \citep{Phi10}, as well as diffuse clouds on the line-of-sight toward W49N, W51 \citep{Son10}, W31C \citep{Neu10}, Sgr B2 (M) \citep{Mon11a} and in some nearby active galaxies, such as, Mrk 231 \citep{van10} and Arp 220 \citep{Ran11}. }
%\textcolor{red}{The HF $J=1-0$ transition is generally observed in absorption, as expected, due to its very large {\it A} coefficient. Only an extremely dense region or a strong radiation field could generate enough excitation to yield an HF feature with a positive frequency-integrated flux. }
\cite{van12} present HF observations toward the Orion Bar,  where the line appears in emission and demonstrate that the excitation of HF is dominated by collisions with electrons, and argue that similar conditions can be found toward active galaxies such as Mrk 231 \citep{van10}, where the radiation field is strong and the HF line also appears in emission.

The HIFI observations support the theoretical predictions that HF will be the dominant reservoir of interstellar fluorine \emph{under a wide range of conditions}. The most interesting aspect is that HF will likely be an important tracer of molecular gas in high-redshift galaxies, as seen by recent observations with the Caltech Submillimeter Observatory (CSO) that revealed the highest-redshift detection of interstellar HF to date toward the luminous lensed Cloverleaf galaxy at {\it z} = 2.558 \citep{Mon11b}. High-redshift HF observations will be done routinely with the large collecting area of the Atacama Large Millimeter/submillimeter Array (ALMA). However, to qualitatively analyze the distant galaxy measurements, a good understanding of HF in the galaxies in the local universe is needed, which to date, is only achievable with {\it Herschel}. In this paper, we study the HF content of two nearby galaxies NGC~253 and NGC~4945, prototypical examples of a nearby starburst and a composite active galactic nucleus (AGN) -- starburst nuclei, respectively. NGC~253 is an edge-on ({\it i} = 78$^{\circ}$) nearby barred galaxy associated with the Sculptor group at a distance of {\it D}~=~3.5~Mpc, having an IR luminosity of {\it L}$_{FIR}$ = 1.7 $\times$ 10$^{10}${\it L}$_{\odot}$ \citep{Rad01} originated in intense massive star formation regions within its central few hundred parsecs \citep{Str04}. It is also suggested that NGC~253 contains a weak AGN in conjunction with the strong starburst \citep[e.g.][]{Mul10}. NGC~4945 is also a nearly edge-on ({\it i} = 78$^{\circ}$) local galaxy, part of the Centaurus group at a distance of $\sim$~3.8 Mpc \citep{Kar07}, with a total infrared luminosity of {\it L}$_{FIR}$ = 2.4 $\times$ 10$^{10}${\it L}$_{\odot}$ \citep{Bro88}. %NGC~4945 host both an AGN and a starburst. It is believed to be a post starburst phase (koornneef 1993) or in a transition from a starburst to a Seyfert galaxy (Moorwood \& Oliva 1994). \textcolor{red}{more...}

In Sections 2 and 3 the observations and results are described. In Section 3.1 we obtain the HF column densities for each galaxy. Outflow and inflow gas properties and their nature are discussed in Section 3.2 and 3.3, with conclusions in Section 4.

\section{Observations}

Using the {\it Herschel}/HIFI band 5a receiver, we observe the $J=1-0$ line of HF toward NGC~253 ($\alpha$$_{J2000}$ $=$ 00$^h$47$^m$33.1$^s$ and $\delta$$\rm_{J2000}$~$=$~-15$^\circ$17$\arcmin$17.6$\arcsec$) and NGC~4945 ($\alpha$$_{J2000}$ $=$ 13$^h$05$^m$27.48$^s$ and $\delta$$\rm_{J2000}$~$=$~-49$^\circ$28$\arcmin$05.6$\arcsec$).  The rest frequency of the line is 1232.4762 GHz \citep{Nol87}. The NGC~253 data was observed as part of the OT2 (Open Time 2) proposal: {\it Hydrogen Fluoride Absorption Toward Luminous Infrared Galaxies} (PI: S. Lord), while the NGC~4945 data was obtained as part of the {\it Herschel} EXtraGALactic (HEXGAL) key project (PI: R. G\"{u}sten). Observations were made using the dual beam switch mode with reference beams located at PA 260 and 296 degrees for NGC~253 and NGC~4945, respectively.
%\textcolor{red}{The dual beam switch observing mode was used with reference beams located at PA 260 and 296 degrees for NGC~253 and NGC~4945, respectively.}
We used the HIFI Wide Band Spectrometer (WBS), providing a spectral resolution of 1.1~MHz, corresponding to a velocity resolution of 0.27 km~s$^{-1}$ at the frequency of $J =1-0$ line, over a 4 GHz IF bandwidth. 
The observations of HF towards NGC~253 were obtained using three nearly adjacent local oscillator (LO) settings that were average to produce the final spectra.

The initial data reduction was carried out using the {\it Herschel} interactive processing environment (HIPE) (Ott et al. 2010) with pipeline version 9.
%The data have been reduced using the {\it Herschel} interactive processing environment (HIPE) (Ott et al. 2010) with pipeline version 9. 
 The IRAM GILDAS package\footnote{http://www.iram.fr/IRAMFR/GILDAS/} was then used to average the individual spectra scans and for the subsequent data analysis.
%\textcolor{red}{The resulting Level 2 double sideband (DSB) spectra were exported to the FITS format for a subsequent data reduction and analysis using the IRAM GILDAS package\footnote{http://www.iram.fr/IRAMFR/GILDAS/}. }
The HIFI beam size at the line frequency is 17$\arcsec$ with assumed main beam efficiency ($\eta$$\rm_{mb}$) of 0.64 \citep{Roe12}. 
The resultant spectra present a DSB continuum main beam temperature of 1.14 and 1.66 K, and a rms noise of 22 and 36 mK at a velocity resolution of 2 km~s$^{-1}$ toward NGC~253 and NGC~4945, respectively. The data quality is excellent as shown in Figure 1, where a zeroth order baseline is indicated with the horizontal solid line. 
 
% \textcolor{green}{It is important to address what motivated the choice of a linear baseline and if there is any indication that the baseline could be of a higher order polynomial as this might mask the emission peak seen in NGC 253 and affect its column density.}

\section{Results}
 Figure 1 shows the HF $J=1-0$ line observed toward the nuclei of NGC~253 and NGC~4945. The line spectrum toward NGC~253 shows blue shifted absorption and redshifted emission, i.e. a P--Cygni profile. The high resolution absorption HF line spectrum toward NGC~253 covers a velocity range of about $\sim$~80~--~283~km~s$^{-1}$, and reveals a shift in velocity with respect to the central velocity. The absorption feature is centered at $\sim$~196~km~s$^{-1}$ a much lower velocity than the systemic velocity, 235 km~s$^{-1}$ marked with the dashed line in Figure~1, with the deepest absorption feature even further blue shifted. The HF spectrum is very similar to the H \textsc{i} absorption profile (Figure~2) observed toward the nucleus of NGC~253 \citep{Kor95} and several OH spectral lines \citep{Tur85,Bra99, Stu11} covering a similar velocity range to the HF absorption line and also showing an equivalent velocity shift. 

The HF spectrum toward NGC~4945 shows an asymmetric absorption profile, with at least two velocity components, in the velocity range of $\sim$~460~--~715~km~s$^{-1}$. To isolate  each individual velocity component, and obtain the corresponding line width and center velocity, we fit the spectrum with two Gaussian components centered  at $\sim$~540 and 640~km~s$^{-1}$ (see Figure~3). These two absorption features are also observed in the H \textsc{i} absorption spectrum as shown in Figure~3.

%The spectra line toward NGC~4945 is only showed in absorption. The line widths can help to determine the nature of the absorbing systems. Larger line widths greater than 100 km~s$^{-1}$, are consistent with a disturbed interacting galaxy, and inconsistent with single narrow line absorbing clouds like those within the disk of the Milky Way. 
%, and a second one shifted by $\sim$~50~km~s$^{-1}$ from the systemic velocity of the galaxy, V$_{SYS}$ $\sim$ 585 km~s$^{-1}$ \citep{Cho07a}}. 
%, centered close to the systematic velocity of $\sim$~560 km~s$^{-1}$ (dashed line in Figure 1)

\subsection{Column Densities of the Absorbing HF gas}
From the line-to-continuum ratio in the absorption profiles we can obtain a direct measurement of the HF column densities. First, we derive apparent optical depths of the HF lines ($\tau$ $=$ {\it -ln}[2{\it T$\rm_L$/T$\rm_C$} -- 1], where the spectrum is normalized with respect to the single sideband continuum), assuming that the foreground absorption covers the continuum source\footnote{The continuum at 180 $\mu$m and 350 $\mu$m extents to a diameter of $\sim$~10~kpc  \citep{Mel02}  and $\sim$~1.6~kpc \citep{Gea86}, respectively, much larger than the HIFI beam at the HF wavelength, equivalent to a 0.3~kpc diameter on the source.} entirely, and that all HF molecules are in the ground rotational state, due to the large spontaneous emission coefficient of HF and low rates of collisional excitation. The resulting normalized spectra are shown in Figure~2 \& 3. For NGC~253, we calculate the optical depth from the absorption profile obtained after subtracting a Gaussian line centered at the systemic velocity fitting the emission component of the HF spectrum (see Figure~2). The velocity integrated optical depth ($\int$$\tau${\it d}V) over the velocity interval from 63 to 295 km~s$^{-1}$ for NGC~253 and from  445 and 720 km~s$^{-1}$ for NGC~4945, is 100 and 115 km~s$^{-1}$, respectively. We thus derive the HF column densities for the correspondent LSR velocity range using Equation (3) of \cite{Neu10} and obtain total HF column densities of 2.41 $\pm$ 0.73 $\times$ 10$^{14}$ and 2.77 $\pm$ 0.85  $\times$ 10$^{14}$ cm$^{-2}$ toward NGC~253 and NGC~4945, respectively. %Using the CO data from \cite{Mau96a} and \cite{Mau96a}, and a Galactic {\it N}(H$_2$)/{\it I}(CO) conversion factor of 2.5 $\times$ 10$^{20}$ cm$^{-2}$/K km~s$^{-1}$, we estimate the H$_2$ column densities, {\it N}(H$_2$), shown in Table 1. Since the HF line is seen in absorption, the relevant H$_2$ column density, for the HF abundance calculation, is that in front of the continuum source, which should be $\sim$ 1/2 of the total hydrogen column density. Thus, we obtain HF abundances with respect to H$_2$ of 1.13 $\pm$ 0.43 $\times$ 10$^{-9}$ and 2.44  $\pm$ 0.75 $\times$ 10$^{-9}$ for NGC~253 and NGC~4945, respectively. 
The uncertainties in the resultant column densities are originated from the random noise and the systematic errors introduced by the calibration uncertainties \citep[][]{Roe12}

\subsection{Gas Kinematics~--~Molecular Outflow in NGC~253}

%\subsection{Gas Kinematics}

A rotating nuclear disk of cold gas in NGC~253 was suggested as a result of the velocity shift of the absorption feature seen in the H \textsc{i} spectrum \citep{Kor95}. In addition to the rotation, the P--Cygni profile present in the HF spectrum line and also in the OH spectra \citep[e.g.,][]{Stu11} indicates a radial motion of gas, characteristic of an outflow. The blue--shifted absorption must arise from a region in front of the continuum originating from the nucleus and moving away from it, and toward us along the line of sight. The redshifted emission in the HF profile, must then originate from gas behind the continuum source and moving away from the nucleus and us. To estimate the molecular mass in the outflow, we adopt an HF abundance relative to H$_2$ of 3.6 $\times$ 10$^{-8}$, based on the chemical predictions of \cite{Neu05}, consistent also with the HF abundances observed toward galactic diffuse clouds \citep{Son10,Mon11a} and outflows \citep[e.g.][]{Emp12}. In high-mass star-forming regions though, where the densities are much higher compare to those of diffuse clouds, the obtained HF abundances are lower by about two order of magnitude (e.g., \citealt{Emp12} derived HF abundances as low as 5~$\times$~10$^{-10}$ in the denser parts of the massive star formation region of NGC 6334 I). The low abundance of HF under these dense (10$^5$~-~10$^6$~cm$^{-3}$) and warm (100~-~150 K) conditions is most likely caused by freeze-out of HF onto dust grains. The polar nature of HF implies large desorption energy. However, dynamically active regions such as outflows and inflows can provide the needed energy for this desorption, and HF abundances in these active regions are similar to those found in diffuse clouds \citep[few 10$^{-8}$,][]{Emp12}. This explanation is supported by evidence for thermal desorption of another polar molecule, H$_2$O, in dynamically active outflow \citep{Fra08,Kri10}.
%The low abundance of HF under these conditions is most likely caused by freeze--out 
%\textcolor{red}{of HF molecules onto dust grains. This argument is derived from the fact that almost all fluorine should be bound in HF due to its high proton affinity, and that the desorption energy of HF is assumed to be quite large due to its polar nature. In dense (10$^5$~-~10$^6$~cm$^{-3}$) and warm (100~-~150 K), but dynamically active, regions such as outflows and inflows, the observed HF abundances are comparable to values found in diffuse clouds \citep[few 10$^{-8}$,][]{Emp12}. In the picture of HF freeze--out at high densities, this can be explained by thermal desorption of HF from the ice grains, similar to the desorption of water in outflows .}
Thus, using the theoretical HF/H$_2$ abundance and a source size equal to the beam size 17$\arcsec$ ($\sim$ 300 pc), the mass\footnote{The mass is given by {\it M}$_{H_2}$ $=$ m$_H$ $\times$ $\mu$$_{H_2}$ $\times$ {\it N}$_{H_2}$ $\times$ {\it D}$^2$ $\times$ $\Omega$, where {\it m}$_H$ is the H-atom mass, $\mu$$_{H_2}$ $\sim$ 2.8 is the molecular weight per hydrogen molecule, {\it D} is the distance to the source and $\Omega$ = $\theta$$^2$$\pi$/4 is the solid angle, with $\theta$ the angular diameter.} of the outflowing gas is {\it M}(H$_2$)$_{out}$ $=$ 0.97 $\times$ 10$^7$ M$_\odot$.  
The outflow projected maximum (terminal) velocity measured from the absorbing gas is about 130 km~s$^{-1}$, measured from the velocity interval where the absorption is deeper than 3$\sigma$ ($\nu$$_ {\rm \scriptscriptstyle SYS}$ - 100) km~s$^{-1}$. The outflow in NGC~253 has its central axis normal to the galaxy disk \citep[i.e. {\it i} = 12$^{\circ}$,][]{Wes11}, then the actual maximum outflow velocity is $\sim$ 194 km~s$^{-1}$. The outflow extends from a few hundred pc to tens of kpc, with the majority of the molecular gas located within a radius of $\le$ 1~kpc \citep{You95}. We assume a compact outflow scenario which extends to 300 pc, the region of intense star-formation activity, and the fully extension of the molecular outflow of 1 kpc. Thus, a range of outflow radii from $\sim$ 300 -- 1000 pc imply a dynamical time of {\it t}$\rm_{dyn}$ $=$ {\it R/V} $=$ 1.51 -- 5.03 $\times$ 10$^{6}$ yrs. The resulting molecular mass loss rate is $\delta${\it M}/$\delta${\it t} $=$ {\it \.{M}} $=$ 1.93 -- 6.43~M$_\odot$ yr$^{-1}$. Note that these values are consistent with that obtained using the OH absorption line \citep{Stu11} and the more direct measurements of CO from \cite{Bol13}, which imply values of 1.6$^{+4.8}_{-1.2}$ and $\sim$~3~M$_\odot$~yr$^{-1}$, respectively. The outflow rate estimates from HF provide a measurement independent of uncertainties introduced by the CO-to-H$_2$ conversion factor, $\alpha$$_{\rm \scriptscriptstyle CO}$, and even though the derived {\it \.{M}} depends on the adopted HF abundance relative to H$_2$, lower HF abundances will lead to larger outflow rates.

 %Due to the galaxy inclination ({\it i} $=$ 78) Westmoquette teal. 2010/11 for the reference on the inclination of the outflow for NGC~253
 
 %$^{12}$CO-luminosity to {\it M}(H$_2$) conversion factor\footnote{The  $^{12}$CO-luminosity to {\it M}(H$_2$) conversion factor: {\it M}(H$_2$) $=$ 3.47 $\times$ {\it L}($^{12}$CO), with $^{12}$CO luminosity given by {\it L}($^{12}$CO) $=$ 18.5 $\times$ {\it D}$^2$ $\times$ $\theta$$^2$ $\times$ {\it I}($^{12}$CO) K km~s$^{-1}$ pc$^2$ (where {\it D} is in Mpc, $\theta$ in $\arcsec$ and {\it I}($^{12}$CO) in K km~s$^{-1}$)} and the $^{12}$CO luminosity of \cite{Mau96a}. This results in a mass of the outflowing gas of {\it M}(H$_2$)$_{out}$ $=$ 1.04 $\times$ 10$^8$ M$_\odot$.  
 
 \subsubsection{What mechanism is driving the outflow?}

Galactic-scale outflows are common phenomena, detected in most active star-forming galaxies in the local universe \citep{Col96} and at high redshifts (Pettini et al. 2001). They are the key mechanism by which energy and metals are recycled in galaxies and deposited into the intergalactic medium. Outflows are thus closely connected with galaxy formation and evolution, potentially regulating the growth of massive galaxies and the thermal properties of the intracluster medium in galaxy groups and clusters \citep[e.g.,][]{Sil98}. To date, most of the relevant information on outflows has come from observations of X-ray emission produced by the hot gas, the optical line emission produced by the warm gas, and interstellar absorption lines. Molecular outflows have also been detected by means of $^{12}$CO and high-density tracers -- such as HCN -- line wings \citep[e.g.,][]{Aal12}. The dominant outflow mechanism, whether {\it starburst driven} where the winds are driven by the mechanical energy and momentum from stellar winds and supernovae (SNe); or {\it AGN driven}, where the black hole (BH) activity may also trigger outflows by accretion, is rarely unambiguously determined \citep[see review by][]{Vei05}. In the case of NGC~253, there is extensive observational evidence that points toward the starburst driven outflow scenario in its central region \citep{Str00, For00, Wea02}. The line widths can help to determine the nature of the outflow, as described in \cite{Stu11}, where a coarse correlation between the outflow velocity and {\it L}$_{AGN}$ is obtained. Furthermore, studies of \cite{Rup05} and \cite{Kru10} show that ULIRGs with high AGN fractions have high outflow velocities ($\ge$~1000~km~s$^{-1}$), while the HF absorption line and other molecular lines, such as OH seen toward NGC~253, present line widths of less than 300 km~s$^{-1}$. Assuming an outflow driven by the wind-momentum of massive stars, we estimate the star formation rate \citep[SFR(M$_\odot$ yr$^{-1}$) $\approx$ {\it L$_{IR}$} $\times$ 10$^{-10}$,][]{Ken98} of 1.7 M$_\odot$ yr$^{-1}$, calculated from the IR luminosities (with FIR solely due to star formation). The SFR is comparable to the outflow rates ({\it \.{M}}) calculated in the previous section with the HF outflow parameters, when one takes into account the uncertainties in the outflow mass and structure, resulting from the lack of angular resolution to determine the spatial distribution of the outflow material traced by HF. Thus, it appears that the SFR could expel the cold gas by wind-momentum of massive stars.  %and will have a depletion time \citep[{\it M}$_{gas}$/\.{M}, where {\it M}$_{gas}$ is taken from][]{Gra11} of $\sim$~5 $\times$~10$^{7}$ yr, which is more than an order or magnitude greater than the time calculated in previous section with the HF outflow parameters. 
Another interesting mechanism that can drive the outflow is the radiation pressure from the absorption and scattering of starlight by dust grains. The radiation can be produced both by a starburst or AGN. \cite{Mur05} suggested that for a radiation pressure driven outflow the total momentum deposition rate ({\it \.{P}~$\approx$~ \.{M}V}) should be equivalent to the total momentum flux ({\it L/c}, where {\it L} is the starburst contribution to the total luminosity). For NGC~253, the momentum deposition rate is comparable to the total momentum flux. %However, due to the large uncertainties in the outflow mass and structure, resulting from the lack of angular resolution to determine the spatial distribution of the outflow material traced by HF, 
Therefore, the pressure driven scenario cannot be discarded. %, for instance taking the values (outflow velocity and masses ) from the OH data obtained from \cite{Stu11}, the total momentum flux has equivalent values to the total deposition rate.  

%,

\subsection{Gas Kinematics in NGC~4945}

The HF observations in NGC~4945 show an absorption feature with at least two components centered at $\sim$ 540 and 640~km~s$^{-1}$, shown in Figure~3. For comparison Figure~3 also shows the absorption profile of H \textsc{i} \citep{Ott01} observed towards the nucleus. These almost axi-symmetric gas components at $\nu$$_ {\rm \scriptscriptstyle SYS}$ $\pm$ 50~km~s$^{-1}$, taking $\nu$$_ {\rm \scriptscriptstyle SYS}$ $\sim$ 585~km~s$^{-1}$ \citep{Cho07}, can indicate two possible scenarios, the presence of a nuclear {\it molecular gas ring} and an {\it inflowing gas motion}. The existence of rapidly rotating clouds surrounding the nucleus has been suggested previously in studies of several molecular lines, such as OH \citep{WW90}, CO \citep{Whi90} and H \textsc{i} \citep{Ott01} line observations, where spatially resolved maps of the nucleus of NGC~4945 support the existence of two concentric fast-rotating nuclear rings surrounding the nucleus within a radius of $\le$~200~pc.
%From the CO position-velocity maps from \cite{Dah93}, the 640 km~s$^{-1}$ velocity component is only seen inside the gas ring at {\it R} $\le$ 200 pc supporting the scenario of two concentric rotating cloud rings. 
%Both velocity components from the HF spectrum are also observed in the H \textsc{i} absorption spectrum (see Figure 3) and the OH line \citep[see Figure 3 in][]{Whi90}, where the feature at 640 km~s$^{-1}$ corresponds to the strongest absorption and appear as dips in the central $^{12}$CO $J=2-1$ emission line \citep[see Figure 5 in][]{Dah93}. }
 %\textcolor{black}{. Other spectral lines such us the OH 1667 MHz (REF WHITEOAK AND GARNER, 1975) shows similar line profile with a shift of their center velocity, which were interpret as evidence of a fast-rotating nuclear ring.  } 
Simultaneously, Figure~3 shows a prominent HF and H \textsc{i} feature at 640 km~s$^{-1}$ also observed as an absorption dip in the CO spectra \citep{Dah93} that traces red-shifted foreground gas moving toward the center of the galaxy and can be interpreted as an infall signature. Signatures for inflowing molecular gas at similar velocities have also been found in high-density gas tracers such as, HCN, HCO$^+$ and CN \citep{Hen90}, where an absorption dip is seen in the emission line profiles. 
%The 540 km~s$^{-1}$ ($\nu$$_ {\rm \scriptscriptstyle SYS}$ -- 50 km~s$^{-1}$) component has been seen in emission in most of the other high density molecular gas tracers, although weaker than other velocity components, in most cases \citep{Hen90}. This can be interpreted as an absorption and emission combination caused by a low density gas intervening cloud on the line of sight to the continuum source. %Both velocity components are also observed in the H \textsc{i} absorption spectrum (see Figure 3), where the feature at 640 km~s$^{-1}$ corresponds to the strongest absorption. From the CO position-velocity maps from \cite{Dah93}, the 640 km~s$^{-1}$ velocity component is only seen inside the gas ring at {\it R} $\le$ 200 pc.

Following the discussion in the previous section, we can estimate the inflowing gas physical properties seen toward NGC~4945. The HF column density of the inflowing gas, i. e. within  the velocity interval of 560 to 720~km~s$^{-1}$, is 1.45~$\times$~10$^{14}$~cm$^{-2}$. Assuming an HF abundance with respect to H$_2$ of 3.6~$\times$~10$^{-8}$ and a source size of 8$\arcsec$ ($\approx$ 200~pc), the inflow mass is 1.54~$\times$~10$^{6}$~M$_\odot$. Taking 200~pc as an upper limit to the inflow radius and a maximum inflow velocity of 152~km/s\footnote{ The maximum inflow velocity is estimated using a maximum inflow projected velocity of 134~km~s$^{-1}$ with a nuclear inclination of $\sim$ 62$^{\circ}$, \cite{Cho07}, smaller than the inclination of the large-scale galactic disk of $\sim$ 78$^{\circ}$.},  the derived molecular mass loss rate is {\it \.{M}} $\le$ 1.2~M$_\odot$~yr$^{-1}$. This relative high inflow rate is $\sim$ 3 orders of magnitude greater than that needed to power the AGN itself (accretion rate of 0.0031~M$_\odot$~yr$^{-1}$, \citealt{Lin11}). Evidence of starburst in a region in NGC~4945 of a few parsecs in size, has been shown by \cite{Cho07} using the Submillimeter Array (SMA), where several molecular lines were observed at an angular resolution of a few arcseconds, tracing an inclined rotating disk with the major axis aligned with that of the stardust ring (a 100 pc diameter ring observed in Pa$\alpha$, \citealt{Mar00}). Therefore, this could be a case where the gas inflow is fueling the nuclear starburst. Similar scenarios of inflow fueling the central tens of parsecs starburst have also been reported in analogous galaxies, such as, NGC~1097 \citep{Dav09}.

\section{Conclusions}

We present the detection of the HF $J=1-0$ transition in two nearby galaxies, NGC~253 and NGC~4945. The HF spectrum toward NGC~253 shows a P--Cygni profile, with the absorption blue-shifted with respect to the systemic velocity, as expected for an outflow system. Estimated outflow parameters derived from the HF spectrum are mass of the outflowing gas {\it M}(H$_2$)$_{out}$ $\sim$ 1 $\times$ 10$^7$ M$_\odot$; the molecular mass loss rate {\it \.{M}} $\le$ 6.4 M$_\odot$ yr$^{-1}$; and a maximum unprojected outflow velocity of {\it $\nu$}$\rm_{out}$ $\sim$ 194~km~s$^{-1}$. The relatively small outflow velocity and the SFR comparable to the outflow rate, argue for a starburst nature of the outflow, however other driving mechanisms, such as radiation pressure, cannot be ruled out. While the HF spectrum shows a clear evidence of a molecular outflow, the conclusions about its nature need further confirmation with spatially-resolved observations or a larger statistical sample. The HF line toward NGC~4945 shows an asymmetric absorption feature, where we can distinguish two components at almost axi-symmetric velocities with respect to the systemic velocity ($\nu$$_ {\rm \scriptscriptstyle SYS}$ $\pm$ 50 km~s$^{-1}$). The axisymmetry velocity profile is compatible with the ring and inflowing gas scenarios, where the redshifted component can be interpreted as a signature of gas falling onto the nucleus. The relatively high inflow rate obtained from the absorption spectrum, indicates gas triggering a nuclear starburst. However, higher angular resolution are needed to reach a firmer conclusion about the nuclear kinematics of NGC~4945.  

From these results, the HF $J=1-0$ line promises to provide a valuable probe of the kinematics of absorbing material along the sight-line to bright extragalactic continuum sources. HF $J=1-0$ line observations towards a larger sample of nearby (U)LIRGs, including radiative transfer models will be presented in a future publication. Studies with larger statistical samples will be able to establish a correlation between the HF abundance and the physical conditions, especially density and temperature. The characterization of HF line observations towards nearby galaxies will be an important tool to understand future observations of HF at high redshift, where the HF line could be a relevant probe of molecular gas in dense but especially in lower density regions of those high redshifted sources where the redshifted H \textsc{i} frequency falls outside the frequency allocation windows for radio astronomy.

\acknowledgments

R. M. would like to thank Nick Scoville for reading through the manuscript and providing very helpful comments. HIFI has been designed and built by a consortium of institutes and university departments from across Europe, Canada and the United States under the leadership of SRON Netherlands Institute for Space Research, Groningen, The Netherlands and with major contributions from Germany, France and the US. Consortium members are: Canada: CSA, U.Waterloo; France: CESR, LAB, LERMA, IRAM; Germany: KOSMA,MPIfR, MPS; Ireland, NUI Maynooth; Italy: ASI, IFSI-INAF, Osservatorio Astrofisico di Arcetri-INAF; Netherlands: SRON, TUD; Poland: CAMK, CBK; Spain: Observatorio Astronomico Nacional (IGN), Centro de Astrobiología (CSIC-INTA). Sweden: Chalmers University of Technology-MC2, RSS \& GARD; Onsala Space Observatory; Swedish National Space Board, Stockholm University - Stockholm Observatory; Switzerland: ETH Zurich, FHNW; USA: Caltech, JPL, NHSC. Support for this work was provided by NASA through an award issued by JPL/Caltech.

{\it Facilities:} \facility{{\it Herschel}/HIFI}.

%\clearpage
%\begin{center}
%\begin{deluxetable}{ccccc}
%\tabletypesize{\scriptsize}
%\rotate
%\tablecaption{HF column densities and abundances}
%\tablewidth{285pt}
%\tablehead{
%Source &$\nu$$_{LSR}$& {\it N}(H$_2$) & {\it N}(HF) & X(HF)\\
%&(km~s$^{-1}$)&$\times$~10$^{23}$~(cm$^{-2}$)&$\times$ 10$^{14}$~(cm$^{-2}$)&$\times$ 10$^{-9}$
%}
%\startdata
%NGC~253&235&2.58$^a$ & 1.46 $\pm$ 0.55&1.13 $\pm$ 0.43 \\[2pt]
%NGC~4945&585&2.28$^b$&2.77 $\pm$ 0.85&2.44 $\pm$ 0.75
%	\enddata \\
%\tablenotetext{a}{}
%$^a$The N(H$_2$) is estimated from the CO data of \cite{Mau96a} and $^b$ \cite{Mau96b} 
%\end{deluxetable}
%\end{center}
\clearpage

\begin{figure}
\begin{center}
\begin{tabular}{cc}
\includegraphics[angle=0,scale=.30]{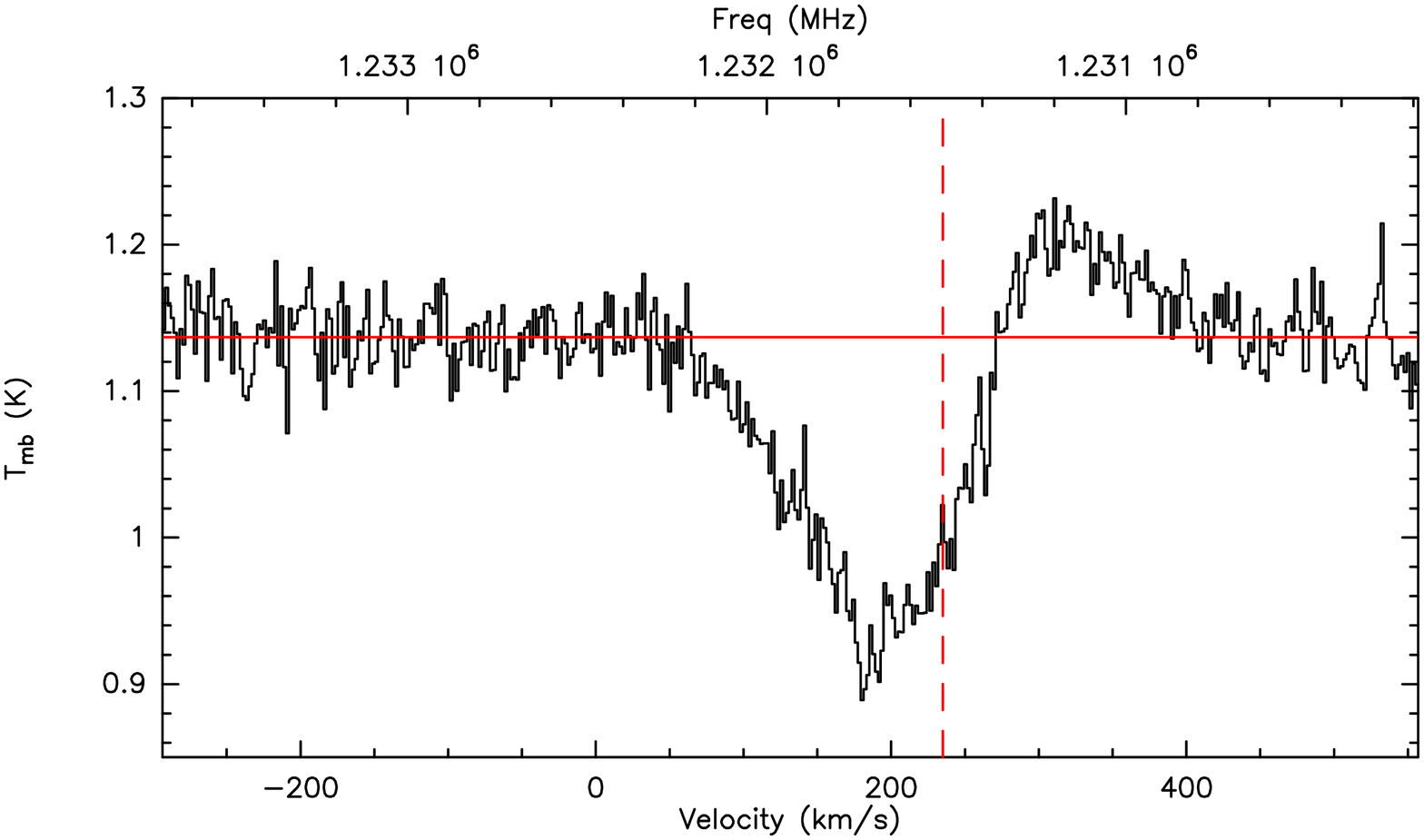}&
\includegraphics[angle=0,scale=.30]{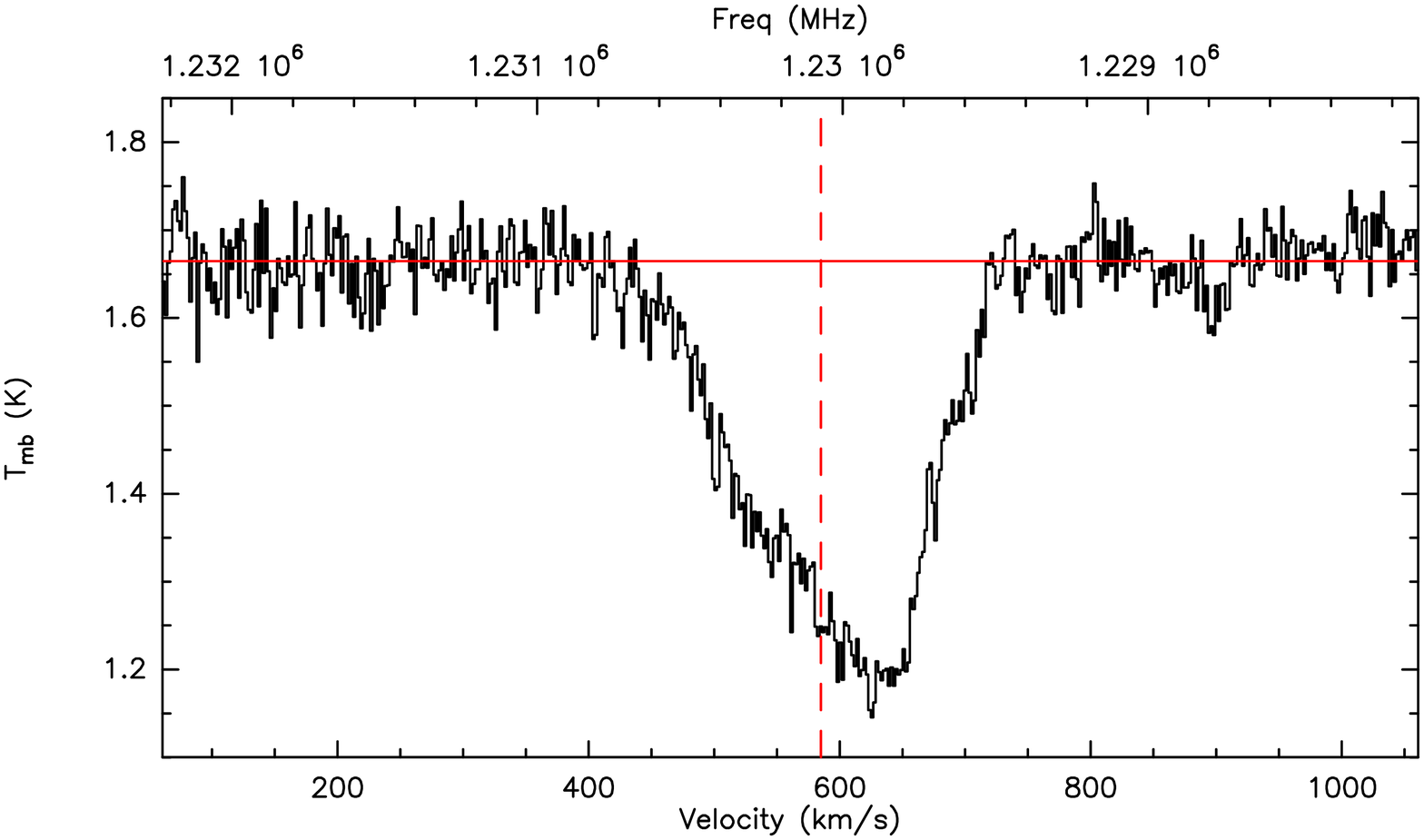}\\
\end{tabular}
\end{center} 
\caption{Spectra of the $J=1-0$ transition of HF toward NGC253 (Left) and NGC~4945 (Right). The velocity scale on the lower axis is with respect to the HF $J=1-0$ rest frequency (1232.4762 GHz). Upper axis shows the frequency scale in MHz. The vertical dashed line corresponds to the systemic $\nu$$\rm_{\rm \scriptscriptstyle SYS}$, the horizontal solid line shows a zeroth order baseline. \label{fig1}}
\end{figure}

%\clearpage

%\begin{center}
%\begin{tabular}{cc}
%\includegraphics[angle=0,scale=.30]{spec_tau_ngc253.eps}&
%\includegraphics[angle=0,scale=.30]{spec_tau_ngc4945.eps}\\
%\end{tabular}
%\end{center} {Fig. 2. Upper panel: normalized spectra of the ground-state transition of HF $J=1-0$ normalized by the corresponding continua toward NGC~253 (Left) and NGC~4945 (Right). The NGC~4945 spectrum also shows the two Gaussian fits center at 543 km~s$^{-1}$ and 637 km~s$^{-1}$ and $\delta$V(FWHM) =  47 km~s$^{-1}$ and $\delta$V(FWHM) =  40 km~s$^{-1}$ , respectively.  Lower panel: optical depth as a function of LSR velocity.\label{fig1}}
%from ~/hf_exgal/ngc253/spec_tau.class
%from ~/hf_exgal/ngc253/spec_tau.class

\clearpage

%\begin{center}
%\begin{tabular}{cc}
%\includegraphics[angle=0,scale=.30]{hi_hf_v2.eps}&
%\includegraphics[angle=0,scale=.30]{hi_hf_ngc4945_v2.eps}\\
%\end{tabular}
%\end{center} {Fig. 2. Upper panel: normalized spectra of the ground-state transition of HF $J=1-0$ normalized by the corresponding continua toward NGC~253 (Left) and NGC~4945 (Right). The NGC~4945 spectrum also shows the two Gaussian fits center at 543 km~s$^{-1}$ and 637 km~s$^{-1}$ and $\delta$V(FWHM) =  47 km~s$^{-1}$ and $\delta$V(FWHM) =  40 km~s$^{-1}$ , respectively.  Lower panel: H I absorption spectrum (in LSR velocities) toward the central continuum source from \cite{Kor95}, the conversion between heliocentric velocities and LSR velocities is $\nu$$_{LSR}$=$\nu$$_{HSR}$ - 2.95 km~s$^{-1}$. The \label{fig1}}

\begin{figure}
\centering
\includegraphics[angle=0,scale=.40]{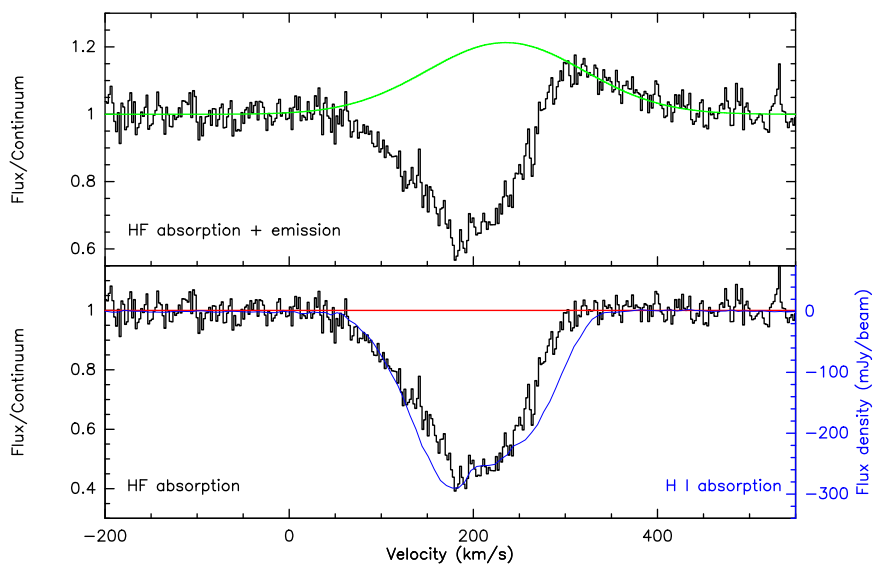}
\caption{Upper panel: spectrum of the ground-state transition of HF $J=1-0$ (black) normalized by the corresponding continuum toward NGC~253 and the Gaussian fit (green) of the emission component. Lower panel: HF normalized absorption profile (black) obtained after subtracting the Gaussian line. H \textsc{i} absorption spectrum (blue, in LSR velocities) toward the central continuum source from \cite{Kor95} is also shown for comparison, the conversion between heliocentric velocities and LSR velocities is $\nu$$_{\rm \scriptscriptstyle LSR}$=$\nu$$_{\rm \scriptscriptstyle HSR}$ -- 2.95 km~s$^{-1}$.  \label{fig2}}
%from ~/hf_exgal/ngc253/spec_tau.class
\end{figure}

\begin{figure}
\centering
\includegraphics[angle=0,scale=.40]{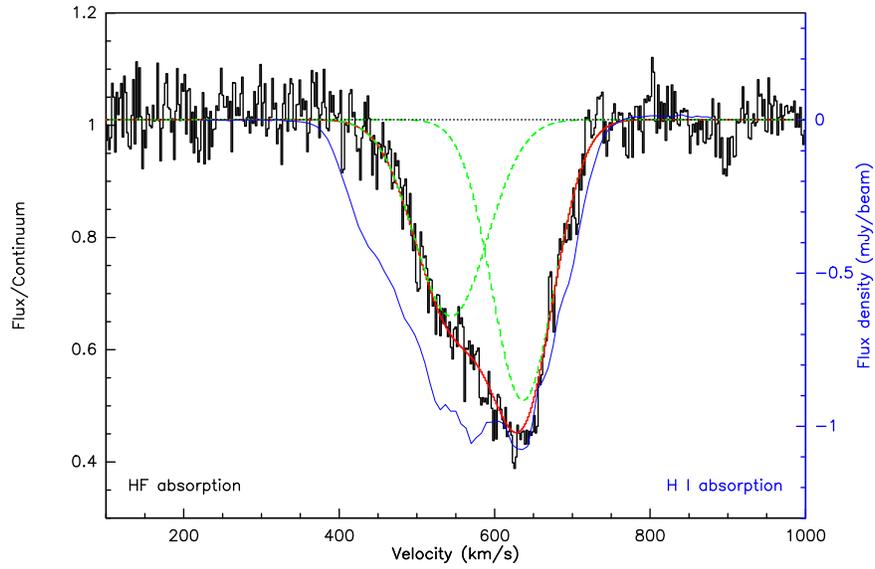}
\caption{Normalized spectrum of the ground-state transition of HF $J=1-0$ (black) with respect to the continuum toward NGC~4945 and two Gaussian fits (dashed green lines) center at 543 km~s$^{-1}$ and 637 km~s$^{-1}$ with $\delta$V(FWHM) =  47 km~s$^{-1}$ and $\delta$V(FWHM) =  40 km~s$^{-1}$, respectively. H \textsc{i} absorption spectrum (blue, in LSR velocities) toward the central continuum source from \cite{Ott01} is shown for comparison.\label{fig3}}
%from ~/hf_exgal/ngc253/spec_tau.class
\end{figure}

%\begin{figure}
%\centering
%\includegraphics[angle=0,scale=.40]{spec_tau_ngc4945.eps}
%\caption{Upper panel: normalized spectrum of the ground-state transition of HF $J=1-0$ with respect to the continuum toward NGC~4945. Lower panel: optical depth as a function of LSR velocity.   \label{fig3}}
%from ~/hf_exgal/ngc253/spec_tau.class
%\end{figure}

\end{document}